\documentclass[3p,times,procedia]{elsarticle}
\usepackage{nupha_ecrc}

\volume{00}

\firstpage{1}

\journalname{Nuclear Physics A}

\runauth{}

\jid{nupha}

\jnltitlelogo{Nuclear Physics A}

\usepackage{amssymb}

\usepackage[figuresright]{rotating}

\begin{document}

\begin{frontmatter}



\dochead{XXVIIIth International Conference on Ultrarelativistic Nucleus-Nucleus Collisions\\ (Quark Matter 2019)}

\title{Jet substructure modification probes the QGP resolution length}


\author[a]{J. Casalderrey-Solana}
\author[b,c]{G. Milhano}
\author[d,e,f]{D. Pablos}
\author[g]{K. Rajagopal}

\address[a]{Departament de F\'\i sica Qu\`antica i Astrof\'\i sica \& Institut de Ci\`encies del Cosmos (ICC), 
\\Universitat de Barcelona, Mart\'{\i}  i Franqu\`es 1, 08028 Barcelona, Spain}
\address[b]{LIP, Av. Prof. Gama Pinto, 2, P-1649-003 Lisboa , Portugal}
\address[c]{Instituto Superior T\'ecnico (IST), Universidade de Lisboa, Av. Rovisco Pais 1, 
1049-001, Lisbon, Portugal}
\address[d]{Department of Physics and Astronomy, Wayne State University, Detroit MI 48201}
\address[e]{Department of Physics and Astronomy, McGill University, Montr\'{e}al QC H3A-2T8}
\address[f]{University of Bergen, Postboks 7803, 5020 Bergen, Norway}
\address[g]{Center for Theoretical Physics, Massachusetts Institute of Technology, 
Cambridge, MA 02139 USA}

\begin{abstract}
We show the sensitivity of groomed jet substructure measurements to the finite resolution power of the QGP to disentangle multiple energetic partons, as produced by QCD jets, that traverse it simultaneously. We illustrate these effects by studying Soft Dropped observables  within a hybrid strong/weak coupling model for jet-medium interactions. By analysing Monte-Carlo-generated jet data in heavy ion collisions, we show that the angular structure
of these type of observables is sensitive to the value of the QGP resolution length. 
 
\end{abstract}


\end{frontmatter}

\def\lres{L_{\rm \small  {res}}}

\def\jraa{R_{\tiny{\rm AA}}^{\rm jet}}
\def\hraa{R_{\tiny{\rm AA}}^{\rm had}}

\def\pTo{{p_{\rm T,1}}}
\def\pTt{{p_{\rm T,2}}}

\def\aSC{{\kappa_{\rm sc}}}

\def\jc#1{{\color{blue}{#1}}}
\def\da#1{{\color{red}{#1}}}
\def\gm#1{{\color{green}{#1}}}

\newcommand{\be}{\begin{equation}}
\newcommand{\ee}{\end{equation}}


\section{Introduction}
\label{intro}
The strong modification of the properties of high energy jets produced in heavy ion collisions serves as a powerful tool with which to study the medium they traverse, namely strongly coupled quark-gluon plasma (QGP). Recent developments on new analysis tools for jet substructure \cite{Andrews:2018jcm} give access to the characterization of relevant features of in-medium parton energy loss.
As we have shown recently in \cite{Casalderrey-Solana:2019ubu}, a hybrid strong/weak coupling model that incorporates finite resolution effects~\cite{Hulcher:2017cpt} can 
reproduce qualitatively and quantitatively the main features of measured jet substructure observables.  
As we stressed in \cite{Casalderrey-Solana:2019ubu}, these observables are very sensitive to the fact that  
 jets in an ensemble that all have a given energy, and that were all reconstructed with a given radius parameter $R$, nevertheless differ wildly from each other due to the stochastic nature of the parton branching process. We showed that the ability of the medium to resolve such differences leaves a strong imprint on measured jet observables, suggesting that these are well suited for the extraction of the QGP resolution length through comparison with experimental data.
 In these proceedings we report additional Monte Carlo studies within our model that  
highlight the role that the QGP resolution length plays in the quenching of multi-partonic configurations. 

\section{The hybrid strong/weak coupling model}
\label{model}

The evolution of an energetic, highly virtual parton is described by the DGLAP equations. Through successive splittings, the high virtuality $Q$ is relaxed down to the hadronization scale. In between splittings, partons in a heavy-ion collision interact with the plasma of temperature $T$. The wide separation between the relevant scales of the system, where $Q \gg T \sim \Lambda_{\rm QCD}$, motivates the implementation of an hybrid description in which the branching process is described perturbatively while the interaction with the QGP is modelled non-perturbatively. Within the hybrid strong/weak coupling model \cite{Casalderrey-Solana:2014bpa,Casalderrey-Solana:2015vaa,Casalderrey-Solana:2016jvj}, the parton showering process, generated through PYTHIA, is embedded into a heavy-ion environment in which the jet propagates through an expanding droplet of QGP liquid. Energy and momentum are transferred into the QGP according to the energy loss rate derived within the gauge/gravity duality for $\mathcal{N}=4$ SYM at infinite coupling \cite{Chesler:2014jva,Chesler:2015nqz}:

\be
\label{CR_rate}
\left. \frac{d E}{dx}\right|_{\rm strongly~coupled}= - \frac{4}{\pi} E_{\rm in} \frac{x^2}{x_{\rm therm}^2} \frac{1}{\sqrt{x_{\rm therm}^2-x^2}}\,,
\ee
where  $x_{\rm therm}= E_{\rm in}^{1/3}/T^{4/3}2\aSC$ is the maximum distance that the parton can travel within the plasma before completely thermalizing, $E_{\rm in}$ is the initial energy of the parton, $T$ the local temperature of the QGP and $\aSC$ is taken as a free parameter that we fitted to data \cite{Casalderrey-Solana:2018wrw}. The wake in the droplet of QGP that is excited by the passage of the jet accounts for the energy and momentum lost by the jet partons, generating through its decay at the freeze-out hypersurface an excess of soft, thermal hadrons with respect to an unperturbed background \cite{Casalderrey-Solana:2016jvj}. Even though the connection between energy loss and the QGP resolution length, $\lres$, is well established at weak coupling \cite{MehtarTani:2010ma}, this is not (yet) the case at strong coupling. In \cite{Hulcher:2017cpt} some of us argued that such length should be proportional to the Debye screening mass $\lambda_D\sim 1/T$. 
One can thus regard $\lres$ as a property of the medium that characterizes the minimum distance between two colored charges such that they engage with the QGP independently.

\section{Results}
\label{results}

Using jet grooming techniques, such as Soft Drop \cite{Dasgupta:2013ihk,Larkoski:2014wba}, we can study the internal jet structure in a well defined fashion through the analysis of its clustering history. After reconstructing a jet using the anti-$k_T$ algorithm, the same jet is reclustered with the angularly ordered C/A algorithm. Then, the first Soft Drop ``splitting'' is the first de-clustering step to satisfy the Soft Drop condition:
\be\label{SoftDropCondition}
z_g > z_{\rm cut} \left(\frac{\Delta R}{R}\right)^\beta \, , \quad z_g \equiv \frac{{\rm min}  \left(\pTo,\, \pTt\right)}{\pTo+\pTt}
\ee
where $\pTo$ and $\pTt$ are the momenta of the leading and subleading branches, respectively, $\Delta R$ is the angular separation between the branches, $R$ is the radius parameter in the anti-$k_T$ algorithm used to reconstruct the jet, and $z_{\rm cut}$ and $\beta$ are the parameters that define the region of phase space of interest. If one iterates this procedure, following the hardest branch, then one obtains the iterated Soft Drop (ISD) procedure \cite{Frye:2017yrw}, which for example allows for a measure of the Soft Drop multiplicity, $n_{\rm SD}$.

\begin{figure}
\centering 
\includegraphics[width=1\textwidth]{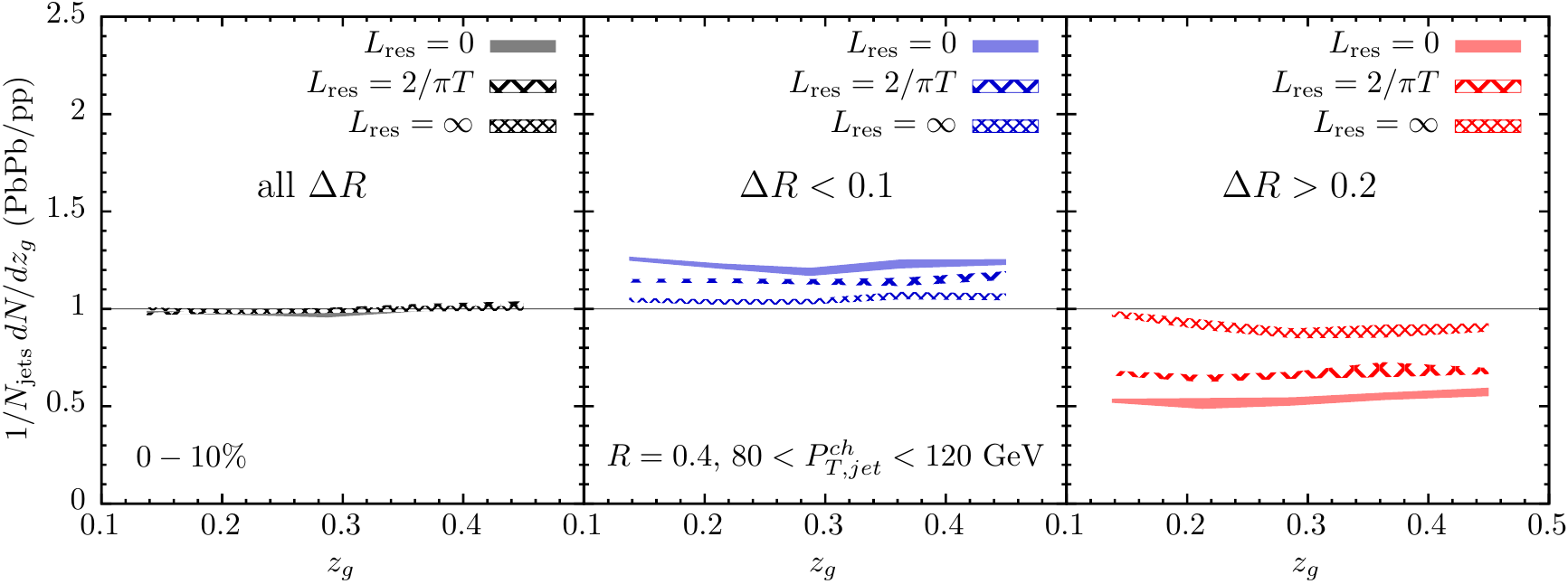}
\caption{\label{Fig:delR} Ratio between predictions for PbPb and pp 
collisions with $\sqrt{S}=2.76$~ATeV of the $z_g$ distribution, inclusive in $\Delta R$ (left panel), for jets with $\Delta R<0.1$ (middle panel) and jets with $\Delta R>0.2$ (right panel), for different values of the QGP resolution length $\lres$. Distributions are normalized to the number of jets entering the analysis, $N_{\rm jets}$.}
\end{figure}

We show in Fig.~\ref{Fig:delR} results for the ratio of the $z_g$ distribution in PbPb and pp collisions obtained using the flat grooming setup, this is with $z_{\rm cut}=0.1$ and $\beta=0$, for the first Soft Drop ``splitting''. We can see that for the results inclusive in $\Delta R$, in the first panel, there is no difference between different choices of $\lres$. Only when we look at the distributions of narrow ($\Delta R<0.1$, middle panel) versus wide ($\Delta R>0.2$, right panel) subjet configurations, we observe a notable separation between the two extreme cases, namely $\lres=0$, where partons are quenched independently from each other from the moment they are formed, and $\lres=\infty$, where a jet is quenched according to its total color charge only, regardless of its internal structure, as if it had never split. In the case with $\lres=0$ where the QGP fully resolves jet substructure, Fig.~\ref{Fig:delR} clearly shows that there is an excess of narrow configurations over wider ones in the final jet sample. The more realistic value $\lres=2/\pi T$ is closer to $\lres=0$ than to $\lres=\infty$. It is worth noting that even though the soft particles from the wake are included in the calculation, their effect in this observable is almost negligible \cite{Casalderrey-Solana:2019ubu} as a consequence of the grooming procedure. (This is not the case for other substructure observables such as the ungroomed jet mass \cite{Casalderrey-Solana:2019ubu} or the $R$ dependence of jet suppression \cite{Pablos:2019ngg}, where the particles coming from the wake formed as the medium responds to the jet play a leading role.)

\begin{figure}   
        \includegraphics[width=0.5\linewidth]{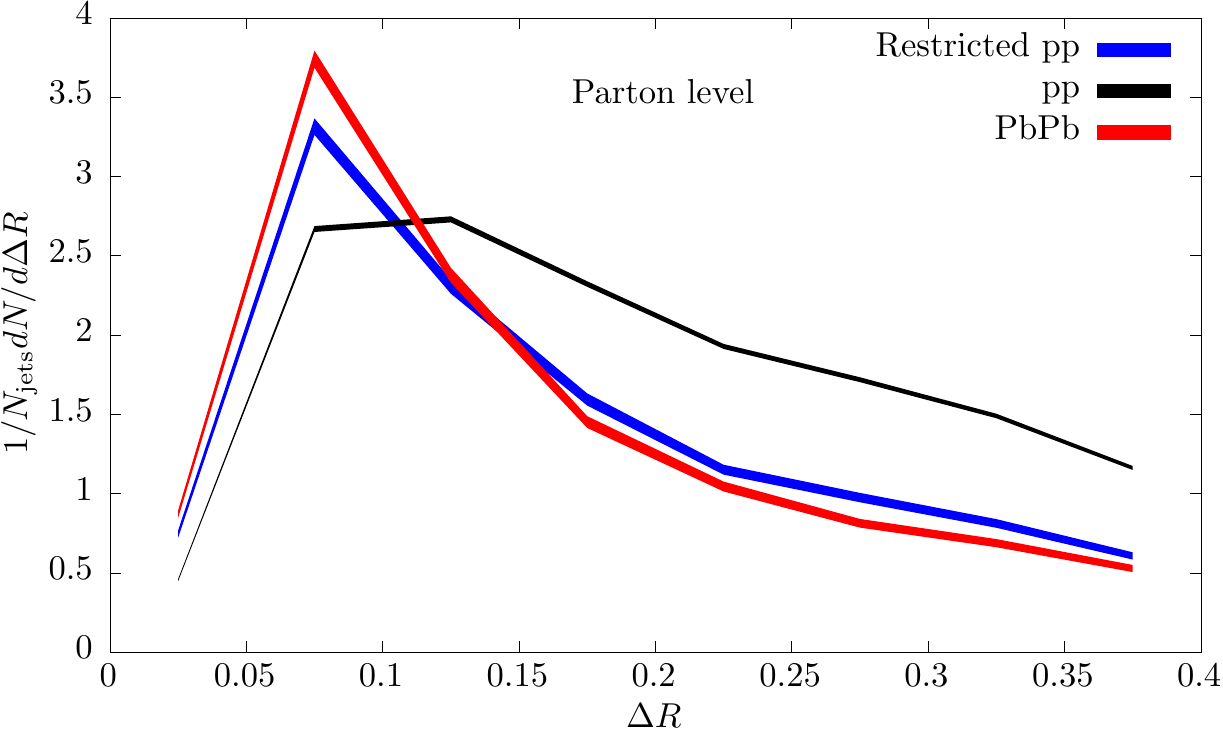} 
        \hfill
        \includegraphics[width=0.5\linewidth]{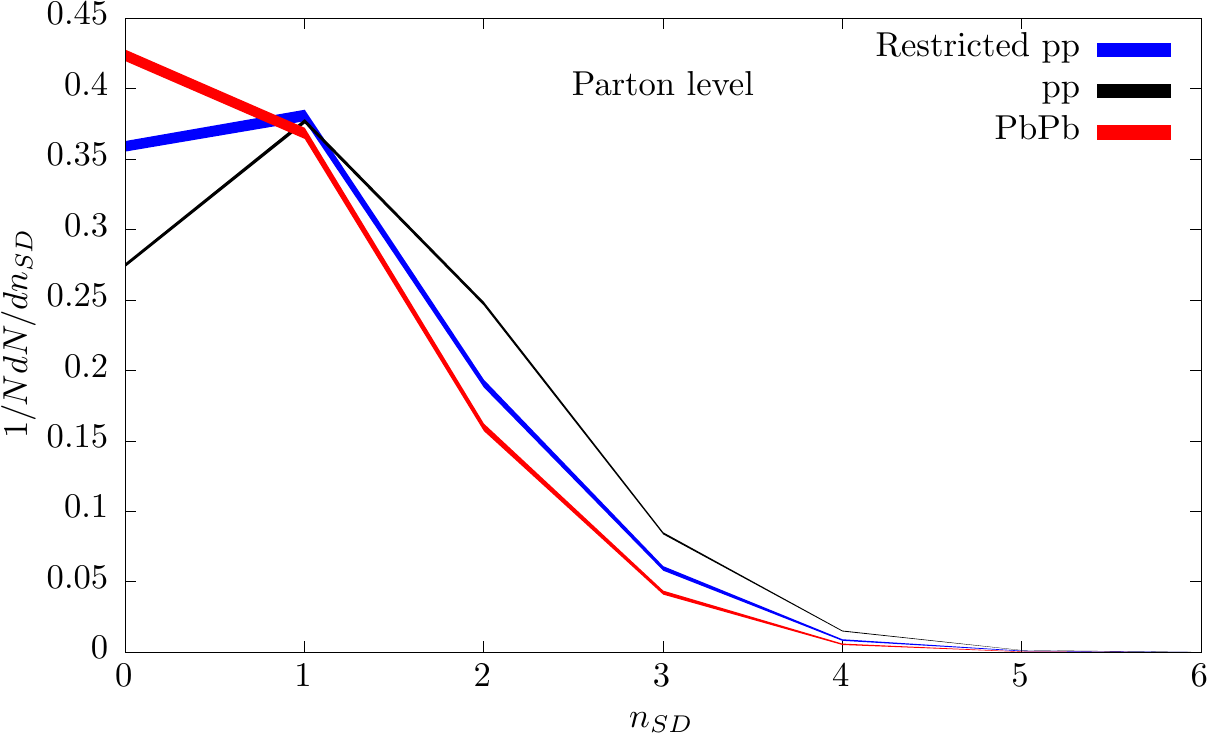} 
        \caption{Parton-level $\Delta R$ (left) and $n_{\rm SD}$ (right) distributions. Black (red) curves are results for pp (PbPb) jets with $p_T>80$ GeV. The blue curves, labelled ``Restricted pp'', are results for those jets that later made it into the PbPb (red) sample, before they were quenched.} \label{Fig:bias}
\end{figure}

In order to better understand the origin of these results, we show in Fig.~\ref{Fig:bias} a set of curves, computed at parton level with $\lres=0$, for the $\Delta R$ (left panel) and the $n_{\rm SD}$ (right panel) distributions. Curves labelled as ``PbPb'' (red) and ``pp'' (black) include all jets with $p_T>80$ GeV. The curves labelled as ``Restricted pp'' include only those vacuum jets which, after quenching, still possess $p_T>80$ GeV; that is, they are pp jets that correspond precisely 
to the ``PbPb sample before quenching''. 
Evidently, this is an analysis that can be done in a Monte Carlo study but only approximately in the analysis of experimental data by following a procedure such as the outlined in \cite{Brewer:2018dfs}.

The message from the left panel is clear: the narrowing of the $\Delta R$ distribution in the PbPb sample is {\it not} due to energy loss reducing the value of $\Delta R$ of individual jets; it {\it is} 
due to a well-known selection bias effect, as follows. Wider jets, those 
which fragment into subjets with a larger $\Delta R$, are more quenched than jets with the same initial $p_T$ that fragment into narrower configurations. 
Since wider jets lose more $p_T$, and since higher $p_T$ jets are much less numerous to 
begin with (the steeply falling jet spectrum), 
wider jets are much less likely to survive in the sample of PbPb jets above a specified momentum cut, when that sample is compared to pp jets selected with the same cut.
The reason why jets with a larger $\Delta R$ in the first Soft Drop splitting are more quenched is because such jets have on average a larger intra-jet activity, thus increasing the number of sources of energy loss \cite{Casalderrey-Solana:2019ubu}. Indeed, as can be seen in the right panel of Fig.~\ref{Fig:bias}, the $n_{\rm SD}$ distribution features a shift towards lower multiplicities which is largely due to the same selection bias effect.

\section{Conclusions}
\label{conclusions}

We have shown that the QGP resolution length, the medium property that controls the degree to which a strongly coupled plasma can resolve the internal structure of a collimated multi-partonic jet shower, has a very strong impact on groomed jet substructure observables, such as the $\Delta R$ dependence of the $z_g$ distribution shown in Fig.~\ref{Fig:delR}. This observable has already been measured by the ALICE collaboration \cite{Acharya:2019djg}; they find that 
after taking into account the effects of background subtraction by smearing our model predictions,
the hybrid strong/weak coupling model with $\lres=0$ can reproduce the experimental results well, including in particular the small $z_g$ enhancement that they find at large $\Delta R$. A low $z_g$ enhancement
has previously been attributed to the medium modification of the splitting function \cite{Chien:2016led,Mehtar-Tani:2016aco,Chang:2017gkt,Caucal:2019uvr} or, alternatively, to the production of hard recoils due to elastic scatterings with quasi-particles from the plasma \cite{Milhano:2017nzm}. 
The fact that our model can reproduce the current data without any actual modification of the splitting function, and with a negligible impact from the soft particles coming from the wake, questions the sensitivity of this observable to the aforementioned effects.  
Looking closely at the ALICE measurements~\cite{Acharya:2019djg}, small discrepancies related to the relative 
contribution from the small and large $\Delta R$ jets suggest that choosing $\lres=2/\pi T$ would yield somewhat better agreement with data. We can also conclude that the completely unresolved case, with $\lres=\infty$, is strongly disfavored by data \cite{Acharya:2019djg}.

The success of this model in predicting the suppression of large $\Delta R$ configurations in favor of narrower ones is due to a selection bias effect that must be present; our analysis shows that it plays a dominant role. 
This effect originates in the combination of a steeply falling jet spectrum with the fact that wider jets with a softer fragmentation pattern lose more energy. In Ref.~\cite{Casalderrey-Solana:2019ubu} we have also checked that, contrary to some prior expectations, the strong ordering in $\Delta R$ does not arise from 
a consideration of when the first splitting occurs, 
but from the fact that the $\Delta R$ of the ``first'' in-cone splitting to pass the soft drop criterion acts as a tagger of the available phase space for future splittings: jets possessing 
subjets separated by a larger $\Delta R$ end up  with a softer fragmentation pattern and, if $\lres$ is finite, hence contain more resolvable sources of energy loss.








\end{document}